\documentclass[reprint,pre,superscriptaddress,floatfix]{revtex4-2}

\usepackage{amsmath}
\usepackage{graphicx,psfrag}
\usepackage[caption=false]{subfig}
\usepackage{verbatim}
\begin{document}
  

\title{Momentum fluctuations in coarse grained systems}


\author{M. Reza Parsa}
\email{mparsa@ucmerced.edu}
\author{Changho Kim}
\email{ckim103@ucmerced.edu}
\affiliation{Department of Applied Mathematics, University of California, Merced, California 95343, USA}
\author{Alexander J. Wagner}
\email{alexander.wagner@ndsu.edu}
\homepage{www.ndsu.edu/pubweb/$\sim$carswagn}
\affiliation{Department of Physics, North Dakota State University, Fargo, North Dakota 58108, USA}


\date{\today}

\begin{abstract}
At first glance the definition of mass and momentum appears to be uniquely defined. We show here, however, that this certainty can be misleading for many coarse grained systems. We show that particularly the fluctuating properties of common definitions of momentum in coarse grained methods like lattice gas and lattice Boltzmann do not agree with a fundamental definition of momentum. In the case of lattice gases, the definition of momentum will even disagree in the limit of large wavelength. For short times we derive analytical representations for the distribution of different momentum measures and  thereby give a full account of these differences.
\end{abstract}


\maketitle

\section{Introduction}
The definition of coarse grained quantities for a mesoscopic simulation method can be surprisingly subtle~\cite{Gorban2006}. We recently investigated fluctuations in coarse grained systems~\cite{parsa2020large}.  We focused there on the behavior of a lattice gas (LG), connected to an underlying molecular dynamics (MD) simulation through the molecular dynamics lattice gas (MDLG) coarse\nobreakdash-graining procedure~\cite{parsa2017lattice}. It had previously been assumed that occupation numbers should follow Poisson statistics. Our research unexpectedly found large fluctuations in the occupation numbers for dense systems. For dilute systems, however, the expectations for ideal gases, namely that the distributions follow a Poisson distribution~\cite{adhikari2005fluctuating}, were indeed verified.

In this article we investigate the connection between the atomistic definition of momentum and the lattice gas (or lattice Boltzmann) definition of momentum~\cite{frisch1986lattice}, in particular the definition of the fluctuating component of the momentum~\cite{boon1991molecular}. It is usually assumed that momentum is a uniquely defined quantity. For coarse-grained descriptions, however, this is not necessarily the case, since space and time averages impact the definition of fluctuating components of the momentum.

We show here that different definitions of microscopic momentum differ significantly. In particular the definition of the fluctuating momentum through a lattice gas deviates significantly from a molecular definition of momentum. The results are highly unexpected: the momentum definitions of a lattice gas differ from molecular dynamics based definitions even in the hydrodynamic limit, i.e. for large wavelength where the usual ideal gas assumptions hold. Even more surprisingly at a larger level of coarse\nobreakdash-graining, where significant correlations are present, MD and LG definitions become identical.

These insights are important when considering the correct implementation of fluctuations in mesoscopic methods like fluctuating lattice Boltzmann~\cite{ladd1993short,adhikari2005fluctuating,dunweg2007statistical}, lattice gas~\cite{grosfils1992spontaneous}, dissipative particle dynamics~\cite{hoogerbrugge1992simulating,espanol1995statistical}, or stochastic rotation dynamics~\cite{ihle2001stochastic,ihle2003stochastic,tuzel2003transport}. We focus here on lattice gas and lattice Boltzmann implementations, and we will use the MDLG procedure to directly map between MD and lattice gas or lattice Boltzmann. In the next section we will briefly introduce lattice gases and the MDLG mapping.

\section{Lattice Gas and MDLG}
The key idea of the MDLG procedure is to map an MD simulation onto a discrete particle evolution on a lattice that has the same formal representation as a lattice gas. To explain this procedure we briefly introduce boolean as well as integer lattice gas models and explain in which key aspects they differ from the MDLG coarse-graining onto a lattice gas. Boolean lattice gases were originally developed as minimal models for statistical mechanics. To simulate hydrodynamic systems, the lattice collisions that conserved both particle number and lattice gas momentum were chosen. It was the key accomplishment by Frisch, Hasslacher and Pomeau~\cite{frisch1986lattice} and Wolfram~\cite{wolfram1986cellular} to select a lattice with sufficient rotational symmetry to recover the Navier-Stokes equations, albeit with some constants that contain unwanted density and velocity dependence. But for practical purposes it was possible to use this method for perfectly adequate fluid simulations, as long as the density variations were not too great. The inherent fluctuations of the lattice gas method were a great advantage for some applications~\cite{ladd1988application,ladd1990dissipative}. It was also of interest how the lattice gas density fluctuations can be related to fundamental molecular theory~\cite{boon1991molecular,rivet2005lattice} or mode coupling theory~\cite{frenkel1989simulation}. However, these fluctuations were not universally helpful, and averaged (i.e. deterministic) lattice Boltzmann methods were developed~\cite{higuera1989boltzmann}. By choosing a different equilibrium distribution it was also possible to remove the density and velocity dependence in the Navier-Stokes equations~\cite{higuera1989lattice,qian1992lattice}. This significant advantage then spurred the development of lattice Boltzmann methods that also included fluctuations~\cite{ladd1994numerical,adhikari2005fluctuating,kaehler2013fluctuating}. It was only recently realized that it is possible to develop a lattice gas method that has the same hydrodynamic limit as an entropic lattice Boltzmann method. This is achieved by allowing for integer occupation numbers, rather than only boolean occupation numbers~\cite{blommel2018integer}. The Boltzmann limit of this lattice gas was shown to be a known entropic lattice Boltzmann method~\cite{ansumali2003minimal}. 

To directly link an MD simulation to a lattice gas we utilize the Molecular Dynamics Lattice Gas procedure~\cite{parsa2017lattice}. It identifies the lattice gas occupation numbers $n_i(x,t)$ as the number of particles that move from lattice  site $\xi-c_i$ at time $t-\Delta t$ to lattice site $\xi$ at time $t$. Here the $c_i$ are lattice displacements, and they are the  set of all distances between lattice sites. For a cubic lattice with lattice spacing $\Delta x$ we get
\begin{equation}
c_{i,\alpha} \in \{0,\pm 1,\pm 2, \cdots \} \frac{\Delta x}{\Delta t}    
\end{equation}
for each Cartesian direction indicated by the Greek symbol $\alpha$.
The $c_i$ are typically referred to as the ``velocity set'' in a lattice gas or lattice Boltzmann context. The only difference here is that the velocity set in principle contains all possible lattice displacements instead of being restricted a priori to a small set of lattice displacements. The effective size of the velocity set then emerges by considering only the lattice displacements for which there are actually particles experiencing them. First we define a function that determines if the coordinates lie inside a lattice site denominated by $\xi$ as
\begin{equation}
    \Delta_\xi(x)=\left\{\begin{array}{ll} 1 &\mbox{ if }x\in \xi\\
    0 & \mbox{ otherwise.}\end{array}\right.
    \label{eqn:Delta}
\end{equation}
We can then express the occupation numbers mathematically as
\begin{equation}
    n_i(\xi,t) = \sum_n \Delta_{\xi-c_i}(x_n(t-\Delta t)) \Delta_\xi(x_n(t)).
    \label{eqn:ni}
\end{equation}
where $x_n(t)$ is the position of the $n$th particle in the MD simulation at time $t$.
Here the $n_i$ take on integer values, and these $n_i$ obey an evolution equation that is formally similar to the standard lattice gas evolution equation:
\begin{equation}
    n_i(\xi+c_i,t+\Delta t) = n_i(\xi,t)+\Xi_i.
\end{equation}
However, it is important to note that in the MDLG context the definition of the $n_i$ is the fundamental quantity given by Eq.~\eqref{eqn:ni}, and the collision operator is determined by the $n_i$ through
\begin{equation}
     \Xi_i = n_i(\xi+c_i,t+\Delta t)-n_i(\xi,t).
\end{equation}
This collision operator is therefore a fluctuating quantity which is fully defined through the underlying MD simulation. The collision operator conserves mass but not momentum. The standard lattice gas definition for the mass and momentum fields are
\begin{align}
    \rho^{LG}(\xi,t)=\sum_i n_i(\xi,t) \label{eqn:rhoLG}\\
    j^{LG}(\xi,t) = \sum_i c_i n_i(\xi,t)  \label{eqn:jLG}
\end{align}
and are also applied here. Note that this definition uses our convention that the particles have a unit mass $m=1$. The average occupation numbers define an equilibrium distribution function
\begin{equation}
    f_i^{eq}=\langle n_i \rangle.
    \label{eqn:feq}
\end{equation}
This equilibrium distribution function is in good agreement of the standard lattice Boltzmann equilibrium distributions only for a very specific choice of lattice spacing $\Delta x$ and timestep size $\Delta t$. 
To achieve this the lattice spacing $\Delta t$ is determined by the constraint
\begin{equation}
    a^2=\frac{\langle \delta  x^2\rangle}{d\Delta x^2} \approx \frac{2}{11}
    \label{eqn:a2}
\end{equation}
where $\delta x=x(t+\Delta t) - x(t)$ is the displacement of a particle during a time interval $\Delta t$, $\langle \delta x^2\rangle$ is the mean squared displacement and $d$ is the number of dimensions. This value of $a^2$ was chosen over $1/6$, which was used in the original publication~\cite{parsa2017lattice}, for purely historical reasons. The tiny difference between the two values for the current study is irrelevant. 
For this choice of $a^2=2/11$ we only have lattice velocities $c_{ix}\in \{-1,0,1\}\Delta x/\Delta t$ with appreciable probabilities~\cite{parsa2017lattice}. 

We know that the underlying momentum in the MD simulation is conserved and that the lattice gas momentum, which only takes on discrete values, cannot be identical to the MD momentum. We therefore want to examine the similarities and differences in these two definitions of momentum.

\section{Definitions for momentum}
For our MD simulations we consider classical particles where the $n$th particle has a position $x_n(t)$ and a velocity $v_n(t)=dx_n(t)/dt$. The density is then simply given by
\begin{equation}
    \rho(x,t) = \sum_n \delta(x-x_n(t))
    \label{eqn:rhodef}
\end{equation}
and the local momentum can be defined as
\begin{equation}
    j(x,t) = \sum_n v_n(t) \delta(x-x_n(t)).
    \label{eqn:jdef}
\end{equation}
To recover continuous density and momentum fields sometimes the Dirac delta function is replaced by a function with a finite base~\cite{espanol1997coarse,rudd1998coarse}, but we do not consider these approaches here. We focus on mesoscale simulation methods that coarse-grain these fields onto a regular lattice (often a square or cubic lattice) with a lattice size of $\Delta x$. 
We can then define the lattice based definition $\rho$ and $j$ as
\begin{align}
    \rho(\xi,t) &= \sum_n \Delta_\xi(x_n(t))\label{eqn:rhoMD}\\
    j(\xi,t) &= \sum_n v_n(t)\Delta_\xi(x_n(t)).
    \label{eqn:jMD}
\end{align}
This is the most straightforward projection of the conserved quantities of mass and momentum onto a lattice.

The definition of the lattice density $\rho^{LG}$ of Eq.~\eqref{eqn:rhoLG} is identical to the definition in the MD context (\ref{eqn:rhoMD})
\begin{equation}
    \rho(\xi,t) = \rho^{LG}(\xi,t).
\end{equation}
As mentioned above the momentum $j$ differs from the lattice gas definition $j^{LG}$. 
In the previous publication~\cite{parsa2017lattice} it was shown that while the instantaneous values of the current $j$ and $j^{LG}$ cannot be identical, the expectation values in the sense of a non\nobreakdash-equilibrium ensemble average are the same:
\begin{equation}
    \langle j(\xi,t)\rangle_{neq} = \langle j^{LG}(\xi,t)\rangle_{neq}
\end{equation}
where $\langle \cdots\rangle_{neq}$ denotes a non-equilibrium  ensemble average.
This has to be the case simply because mass is rigorously conserved both in MD and LG, the definition of mass is identical, and therefore the expectation value of the mass current also has to agree. 

\begin{figure}
    \centering
    \includegraphics[width=\columnwidth]{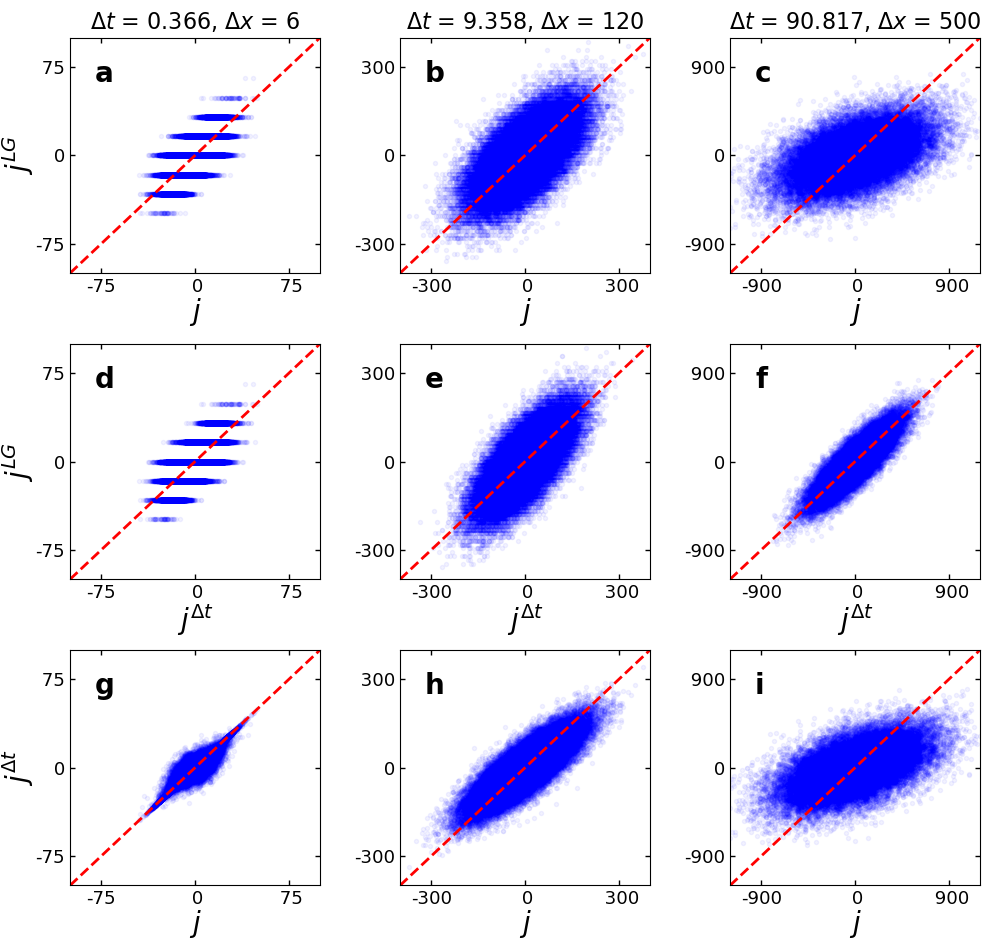}
    \caption{Scatter plots for the correlations of three different momenta considered in this paper for three different time-steps $\Delta t$.}
    \label{fig:scatter}
\end{figure}

It is therefore interesting to examine how well the two non-averaged definitions of the mass current are correlated. To examine the momentum definition we performed equilibrium MD simulations of a two dimensional Lennard-Jones (LJ) fluid given by the inter-particle potential
\begin{equation}
V(x) = 4\epsilon \left[\left(\frac{\sigma}{x}\right)^{12}-\left(\frac{\sigma}{x}\right)^{6}\right].
\end{equation}
For simplicity, we assumed the mass of one particle, $m$, equals one and the unit of time is scaled by $\tau=(m\sigma^2/\epsilon)^{1/2}$. We performed our MD simulations using the LAMMPS package~\cite{plimpton1995fast}. The particles are contained in square box with the length of $3\,000\,\sigma$ with periodic boundary conditions. Uniform configurations of $90\,000$ particles with the kinetic energy corresponding to $50$ in the LJ units were generated as an initial configuration. We ran our simulation with a time step of $0.001\,\tau$. The first $10\,000\,000$ iterations were discarded and then measurements were performed for additional $10\,000\,000$ timesteps. We saved data for specific $\Delta x$ using a time-step that gave $a^2 = 2/11$ from Eq.~\eqref{eqn:a2}. Furthermore, we used ADIOS2 package~\cite{godoy2020adios} along with MD simulations to reduce the time required for reading and analyzing data. For data presented in this paper we used a range of lattice size from $6\,\sigma$ to $500\,\sigma$ which led to different numbers of lattice points for our fixed simulation box.

In Fig.~\ref{fig:scatter} the scatter plot shows the correlation of three different momentum definitions for three different timesteps $\Delta t$. The third definition will be given below, see Eq. (\ref{eqn:jdtdef}).
For $j$ and $j^{LG}$ we can see the correlation in Fig.~\ref{fig:scatter} (a),(b),(c) for three different time-steps.
The two measures $j$ and $j^{LG}$ are indeed not identical,  which would be represented by blue points lying entirely on the red diagonal. Instead we observe noticeable scatter.  Interestingly there is not only scatter, but the average does not even follow the identity, indicating that the widths of the momentum distributions for the two measures are different. Which distribution is wider does depend on the time discretization. One further feature that immediately stands out is the strong discretization of $j^{LG}$ for the shortest timestep. In this case there are on average only $\langle n\rangle= 0.36$ particles in each lattice site.
We can use $f_i^{eq}$ of Eq. (\ref{eqn:feq}) to define the combined occupation density for positive velocities as
\begin{equation}
    f^+=\sum_{i:c_{ix}>0} f_i^{eq} 
    \label{eqn:fplus}
\end{equation}
where the sum is only taking over those indices for which the lattice velocity has a positive $x$-component. Similarly we can define $f^-$ as the sum of all $f_i^{eq}$ where the $c_{ix}<0$.
We also define the average number of particles on each lattice site as
\begin{equation}
    \rho^{eq}=\sum_i f_i^{eq}.
    \label{eqn:rhoeq}
\end{equation}

We can now ask which distribution of particles is expected for lattice momentum $l\Delta x/\Delta t$. If the particles can be considered independent, then the probability that a specific lattice site has the momentum $l\Delta x/\Delta t$ is equivalent to a standard combinatorial problem: given $N$ particles that are to be distributed into three containers. The first container corresponds to a lattice velocity with $c_{ix}=+\Delta x/\Delta t$ at lattice site $\xi$, the second container corresponds to a lattice velocity with $c_{ix}=-\Delta x/\Delta t$ at the same site [the only values entering \eqref{eqn:jLG}], and the last one is the rest of the lattice. This gives the probability of 
\begin{align}
&P\left(j^{LG}_x\equiv l\Delta x/\Delta t\right) 
\nonumber\\
=& \sum_k \frac{N!}{(N-2k-l)!k!(k+l)!}
\nonumber\\& 
\left(1-\frac{f^++f^-}{\rho^{eq}N_\xi}\right)^{N-2k-l}\left(\frac{f^+}{\rho^{eq}N_\xi}\right)^{k+l} \left( \frac{f^-}{\rho^{eq}N_\xi}\right)^k
\label{eqn:jLGdistbin}
\end{align}
to find a total momentum of $l\Delta x/\Delta t$ where $N_\xi$ is the number of lattice sites and we sum over all combinations that have a total momentum of $l \Delta x/\Delta t$, i.e. having $k$ particles with lattice velocity $c_{ix}=-\Delta x/\Delta t$ and $k+l$ particles with lattice velocity $c_{ix}=+\Delta x/\Delta t$. Here we have approximately $f^-\approx \rho^{eq}/6$, $f^0\approx 2\rho^{eq}/3 $, and $f^+\approx \rho^{eq}/6$ where $f^0$ is defined similar to the definition in (\ref{eqn:fplus}) for particles with no $x$ momentum. The number of lattice points is $N_\xi=250,000$. In Fig.~\ref{fig:scatter} we show the values for momenta for $10$ iterations. The expected number of points in the scatter plot with momentum $4\Delta x/\Delta t$ is then $10 N_\xi P(j^{LG}\equiv 4 \Delta x/\Delta t)\approx 1.198$. The expected number of samples with lattice momentum of 5 is  0.014. This is in good agreement with seeing 2 instances of lattice momentum of 4 (as well as none with a lattice momentum of -4), and no instance with a larger absolute lattice momentum. An alternative derivation that does not make the assumption of having only three velocities in any one direction is shown in Section~\ref{subsec:LG}.

This explains why one distribution is discrete while the other is continuous. It does not, however, explain why the momenta are not distributed with the same second moment. To understand this better, we first observe that the momentum distribution $j$ only depends on the instantaneous velocity of the particles, where as $j^{LG}$ depends on the displacement of the particles during the time interval $\Delta t$. We can therefore define an in-between measure consisting of the averaged momentum over the time step $\Delta t$. We follow the same particles that contribute to $j^{LG}$ and time-average their velocities over a period of $\Delta t$:
\begin{align}
    j^{\Delta t}(\xi,t)&= \frac{1}{\Delta t}\sum_n \int_{t-\Delta t}^t dt'\; v_n(t')\Delta_\xi(x_n(t))\\
    &=\sum_n  \frac{x_n(t)-x_n(t-\Delta t)}{\Delta t} \Delta_\xi(x_n(t)).
    \label{eqn:jdtdef}
\end{align}
This is a measure that shares the time step dependence with $j^{LG}$ and at the same time is a measure that only depends on the MD data for the lattice cell. In that sense it is an intermediate measure. For $\Delta t$ much shorter than the mean free time of the particles it should agree with $j$. We also examined the correlation between this measure of the momentum and the fundamental lattice gas momentum in Fig.~\ref{fig:scatter}. Comparing this measure $j^{\Delta t}$ to the fundamental momentum $j$ we see that for the short time of $\Delta t=0.366$ the two measures are indeed highly correlated. But for larger values of $\Delta t$ the variance of $j^{\Delta t}$ becomes smaller than the variance of $j$, and at the same time the scatter increases.

When we compare $j^{\Delta t}$ to $j^{LG}$ we see that for short times the behavior is essentially the same as for $j$ and $j^{LG}$. 
For larger time steps $\Delta t$ the $j^{\Delta t}$ is showing a narrowing of distribution which approaches that of $j^{LG}$. At the same time it can be seen in Fig.~\ref{fig:scatter} that the correlation between $j^{LG}$ and $j^{\Delta t}$ increases with larger $\Delta  t$ and the scatter decreases. It is hard to quantify this effect from a scatter plot, so instead we define a measure for the correlation.

We define a measure for the correlation of two quantities that allows for a relative scaling of the quantities by
\begin{align}
     C_{j^a,j^b}(\Delta t) &= \frac{1}{2XT}\sum_\xi \sum_t \left[\frac{j^{a}(\xi,t)}{\sigma_{j^a}}-\frac{j^{b}(\xi,t)}{\sigma_{j^b}}\right]^2\label{eqn:C12}
\end{align}
where $X$ is the number of lattice sites and $T$  is the number of time-steps we are averaging over. The variance $\sigma_j$ are defined through
\begin{equation}
    \sigma_{j} = \sqrt{\frac{1}{XT}\sum_\xi \sum_t \left[j(\xi,t)-\rho^{eq} u\right]^2}
\end{equation}
where $\rho^{eq} u $ is the average momentum per lattice cell of Eq.~\eqref{eqn:rhoeq}. For our simulations this is set to zero for simplicity.
The measure $C_{j^a,j^b}$ is zero when the two different measures $j^a$ and $j^b$ of momentum are completely correlated (\textit{i.e.} identical up to a scaling factor) and will take on a value of one when they are completely uncorrelated. Note that this measure does take into account that even if the overall scale of the momentum, as given by the spread of the distribution, differs, the momenta could in principle still be strongly correlated. 

\begin{figure}
    \centering
    \includegraphics[width=\columnwidth]{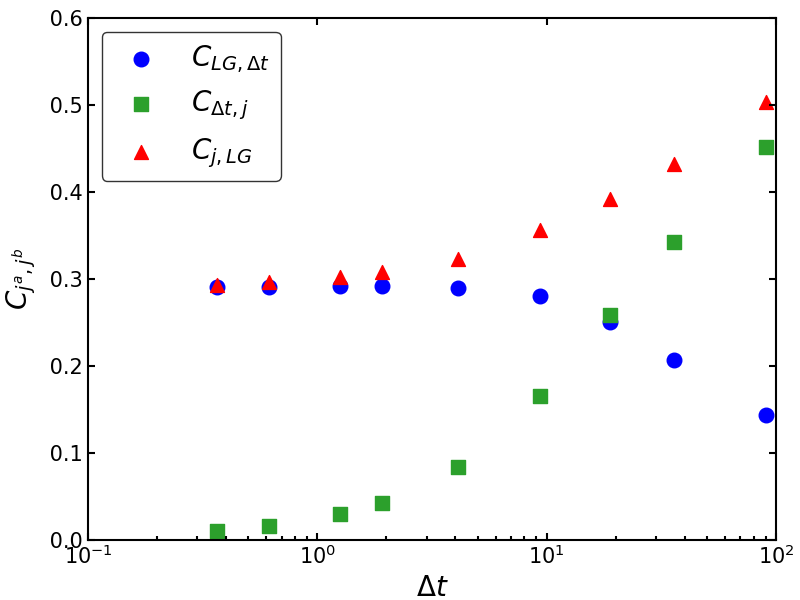}
    \caption{\label{fig:corr} Correlators of Eq.~\eqref{eqn:C12} for the three combinations of the three definitions of momentum considered in this paper. They are shown as a function of the time-step $\Delta t$.}
\end{figure}

The results of measuring these $C_{j^a j^b}$ functions are shown in Fig.~\ref{fig:corr}. As we observed in our discussion of Fig.~\ref{fig:scatter} there is close agreement between the fundamental definition of momentum $j(x,t)$ and the time-averaged momentum $j^{\Delta t}$ for small time intervals $\Delta t$. For larger time steps, this correlation begins to weaken.

We also observe that the correlation between the averaged momentum $j^{\Delta t}$ and the lattice gas momentum $j^{LG}$ does indeed increase with an increasing time-step $\Delta t$ as we surmised from the inspection of Fig.~\ref{fig:scatter}. We hypothesise that this correlation between $j^{\Delta t}$ and $j^{LG}$ will continue to increase. Without being able to give a full proof here, we speculate that at larger coarse-graining the time-averaged velocities that enter both measures become more and more correlated, as was shown in an earlier paper~\cite{parsa2020large}. As coarsening increases, fluctuations become less important and the observed time-averaged velocities will vary less within one lattice cell, leading to agreement between the two measures. Note, however, that this is a subtle effect of time-correlations, and coarse-graining alone is not sufficient, as can be seen in the deterioration of the correlations with the fundamental momentum $j$.

\begin{figure}
    \centering
    \includegraphics[width=\columnwidth]{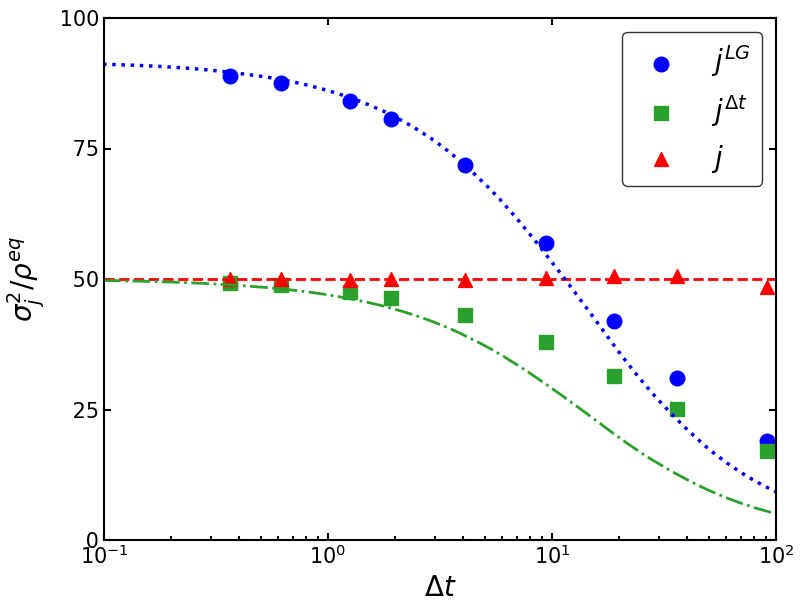}
    \caption{Second moments of the three definitions of momentum from measurement (symbols) compared to theoretical predictions for ideal gas uncorrelated systems of Eqs.~\eqref{eqn:j2},\eqref{eqn:jdt}, and \eqref{eqn:jLGdist} (lines).}
    \label{fig:predict}
\end{figure}

In the above discussion we examined the correlation of the different definitions of momenta normalized by their mean squared displacements. But we have not considered the mean squared displacements themselves. We show those second moments of the momenta, normalized by the average number of particles per cell, in Fig.~\ref{fig:predict} as a function of the time-step $\Delta t$. The most important feature to notice is that the second moment of the lattice gas diverges for small time steps from the results for the instantaneous (and the time\nobreakdash-averaged) momentum. This result was quite unexpected and requires serious consideration.

In the next part of this paper we will consider the distributions of momentum for the three different measures we defined, which can be achieved analytically in the limit of a small time-step. This will elucidate the curious result of the different fluctuation amplitudes for the lattice gas and instantaneous momenta.

\section{Distributions of the momenta \label{sec:dist}}
These findings suggest that there is something unexpected going on with the lattice gas definition of momentum. We will now examine if we can understand these findings quantitatively and how they relate to previous predictions. Firstly let us stress again that an ensemble average of the momentum (which would be free of fluctuations) does agree for all three definitions of the momentum. So we are interested here in the fluctuating component of the momentum. 

\subsection{Instantaneous momentum}
For the lattice momentum measure $j$ the situation is already rather interesting. We can calculate the momentum distribution as
\begin{align}
    P^j(j) =& \int dx_1 dv_1\cdots dx_N dv_N P^N(x_1,v_1,\cdots,x_N,v_N)\nonumber\\
    &\delta\left(j-\sum_{n=1}^N v_n  \Delta_\xi(x_n)\right).
\end{align}
To obtain the average we need the $N$\nobreakdash-particle distribution function in phase space in equilibrium. We assume here that the mean velocity is $u$. We define the average number of particles per lattice cell as
\begin{equation}
    \rho^{eq}= \frac{N(\Delta x)^d}{V}
\end{equation}
which is identical to the definition of Eq.~\eqref{eqn:rhoeq}.
Since we are considering a system in equilibrium here all particles have uncorrelated velocities distributed according to the Maxwell Boltzmann distribution
\begin{equation}
    P^{MB}(v) = \frac{1}{(2\pi k_B T)^{d/2}} \exp\left( -\frac{(v-u)^2}{2k_BT}\right)
\end{equation}
$k_B$ is Boltzmann's constant and $T$ is the temperature. For simplicity we will assume $u=0$ below. The extension to nonzero $u$ is straightforward but makes the expressions below more messy. If we further assume that the particles are dilute enough so that they resemble an ideal gas then the $N$\nobreakdash-particle distribution function is simply given by
\begin{equation}
    P^N(x_1,v_1,\cdots,x_N,v_N,t) = \prod_{n=1}^N \frac{1}{V} P^{MB}(v_n)
    \label{eqn:PNfact}
\end{equation}
otherwise volume exclusion has to be taken into account. In this case the number of particles in a cell will be Poisson distributed and we get
\begin{align}
    &P^j(j)\nonumber\\
    = & \exp(-\rho^{eq})\left[\delta(j)+ \sum_{n=1}^\infty \frac{(\rho^{eq})^n}{n!} \exp\left(-\frac{j^2}{2n k_B T}\right)\right]
    \label{eqn:Pj}
\end{align}
where the surprising appearance of a Dirac delta function represents the finite fraction of lattice cells without particles that contain a momentum of zero. This distribution function is an unusual combination of a discrete set of Gaussians and one $\delta$-function that can be understood as the limit of a Gaussian with zero width. It was the discovery of this unusual distribution function that inspired its application in a quite different context in a recent paper by Pachalieva~\cite{pachalieva2020non}.

Interestingly the second moment of this distribution can be evaluated directly, without requiring the assumption in Eq.~\eqref{eqn:PNfact} of a dilute gas. We get
\begin{align}
    &\langle j(\xi,t)^2\rangle\\
    =&\left\langle \left(\sum_n v_n(t)  \Delta_\xi(x_n(t))\right)^2\right\rangle\\
    =& \int dx_1 dv_1\cdots dx_N dv_N P^N(x_1,v_1,\cdots,x_N,v_N)\nonumber\\
    &\sum_{n=1}^N \sum_{m=1}^N v_n v_m \Delta_\xi(x_n)\Delta_\xi(x_m)\\
   =& N\frac{\Delta x^d}{V} \int dv_1\;(v_1)^2 P^{MB}(v_1)\\
   =& d \rho^{eq} k_B T
   \label{eqn:j2}
\end{align} 
where we have used that the expectation value of a single velocity term is zero by symmetry and that the velocities of any two particles are uncorrelated in equilibrium, leaving on the one\nobreakdash-particle contribution above.
This is a result well known from kinetic theory, and our simulation results are in very good agreement with this theory, as shown in Fig.~\ref{fig:predict}.

\subsection{Displacement momentum}
For the momentum defined from the particle displacement the situation is a little more complicated. Analogously to the previous case we need to calculate 
\begin{align}
    P^{\Delta t}(j) =& \int dx_1 dv_1\cdots dx_N dv_N P^{N,\delta}(x_1,\delta  x_1,\cdots,x_N,\delta x_N)\nonumber\\
    &\delta\left(j-\sum_{n=1}^N \frac{\delta x_n}{\Delta t}  \Delta_\xi(x_n)\right)
\end{align}
but, as was emphasized by Pachalieva \textit{et al.}~\cite{pachalieva2020non}, little is known about the $N$\nobreakdash-particle displacement distribution function. For short time-steps $\Delta t$ it is reasonable to assume that in equilibrium this distribution function will also factorize, but because the time-evolution introduces correlations, this ceases to be valid for larger time-steps. For dilute systems, the buildup of these correlations is less severe~\cite{parsa2020large}, so the assumption of a factorizing distribution function is not immediately invalid.

As a second approximation we will assume here that the factorizing single particle displacement distribution function can be approximated by a Gaussian. We know that this is only approximately true~\cite{pachalieva2020non}, but it will serve here for a rough approximation. We therefore assume
\begin{equation}
    P^{\Delta t}(\delta x) \approx \frac{1}{(2\pi \langle \delta x^2\rangle)^{d/2}} \exp\left( -\frac{(\delta x-u\Delta t)^2}{2\langle \delta x^2\rangle}\right)
    \label{Pdist}
\end{equation}
where the mean squared displacement $\langle \delta x^2\rangle$ is to be obtained from the numerical simulations as usual~\cite{parsa2017lattice}. We also assume that the particles are uniformly distributed. 

With this assumption we can calculate an approximate second moment for the time-averaged momentum:
\begin{align}
    &\langle j^{\Delta t}(\xi,t)^2\rangle\\
    =&\left\langle 
    \left(\sum_n  \frac{x_n(t)-x_n(t-\Delta t)}{\Delta t} \Delta_\xi(x_n(t))\right)^2\right\rangle\\
    =& \int dx_1 d\delta x_1\cdots dx_N d\delta x_N P^N(x_1,\delta x_1,\cdots,x_N,\delta x_N)\nonumber\\
    &\sum_{n=1}^N \sum_{m=1}^N \delta x_n(t) \delta x_m(t) \Delta_\xi(x_n)\Delta_\xi(x_m)\\
   \approx& N\frac{\Delta x^d}{V} \int d\delta x_1\;\left(\frac{(\delta x_1)^2}{\Delta t}\right) P^{\Delta t}(\delta x_1)\\
   =& \rho^{eq} \frac{\langle \delta x^2\rangle}{(\Delta t)^2}.
   \label{eqn:jdt}
\end{align} 
We note that this is only an approximate relation where time-correlations have been neglected. We know a little about the two\nobreakdash-particle displacement distribution from our recent paper~\cite{parsa2020large}, where we showed that correlations in the displacements are decaying exponentially, but this can be very important on the length-scale of a lattice size $\Delta x$. However, it was also shown that this effect is less pronounced for dilute mixtures. So we expect our approximations to be reasonably good for the low-density system considered in this paper as well as shorter time periods $\Delta t$. We see that this is indeed the case in Fig.~\ref{fig:predict}, although deviations from this analytical approximation are visible as soon as the results deviate from the instantaneous momentum. A more detailed analysis of the $N$\nobreakdash-particle displacement distribution function beyond what was presented in~\cite{parsa2020large,pachalieva2020non} will be required to obtain a better analytical approximation, but this is outside the scope of the current paper. 

\subsection{Lattice gas momentum\label{subsec:LG}}
Let us now put our attention on the lattice gas momentum. It was first claimed by Adhikari \textit{et al.}~\cite{adhikari2005fluctuating} and later by D\"unweg \textit{et al.}~\cite{dunweg2007statistical} that the occupation numbers $n_i$ should be Poisson distributed as long as we are considering an ideal gas. This argument was based on the arguments presented in Landau and Lifshitz~\cite{lifschitz1983physical}. A more general derivation of the distribution was given in our previous publication~\cite{parsa2020large}, where it was shown that in general the distribution of the $n_i$ depends on the $N$\nobreakdash-particle distribution function. This is a slight simplification from the derivation of Eq.~\eqref{eqn:jLGdistbin}, where a finite system leading to a multinomial distribution was considered. For that derivation, however, it was assumed that the only values for the velocity in any one direction would be zero or $\pm \Delta x/\Delta t$. The derivation using the assumption of Poisson distributed occupation numbers is simpler than generalizing the previous derivation.

In the special case that this distribution function factorizes, as was assumed in the previous section for the analytical calculation of the momentum distribution for the $j^{\Delta t}$ momentum measure, the occupation numbers $n_i$ are indeed Poisson distributed. With this approximation Parsa \textit{et al.}~\cite{parsa2020large} obtained
\begin{equation}
    P(n_i) = \exp\left(-f_i^{eq}\right) \frac{(f_i^{eq})^{n_i}}{n_i!}
\end{equation}
where the equilibrium distribution are given by Eq.~\eqref{eqn:feq}. A fully analytical expression for $f_i^{eq}$ can be obtained analytically~\cite{parsa2017lattice,parsa2019validity,parsa2018lattice} by solving
\begin{equation}
    f_{i}^{eq}=\int dx \int d\delta  x P^{\Delta t}(\delta x)\Delta_\xi(x)\Delta_{\xi-c_i}(x-\delta x)
\end{equation}
where $\Delta_\xi(x)$ was defined in Eq.~\eqref{eqn:Delta}. While Mathematica is able to obtain an explicit expression for the solution, it is too lengthy to reproduce here. It can be found in the Mathematica notebook in the supplemental material~\cite{supplemental}. 

Of interest here are the first moments of the equilibrium distribution. The zeroth and first velocity moments are given by mass and momentum conservation, as was shown by Parsa \textit{et al.}~\cite{parsa2017lattice}. The second velocity moment, however, is more interesting. It was previously evaluated for $u=0$ in the same publication, but here we also evaluate it for general $u$. The moments are given by
\begin{align}
    \sum_i f_i^{eq} &= \rho^{eq}\label{eqn:f}\\
    \sum_i v_{i\alpha} f_i^{eq} &= \rho^{eq} u_\alpha\label{eqn:fv}\\
    \sum_i v_{i\alpha} v_{i\beta} f_i^{eq}&= \rho u_\alpha u_\beta +\delta_{\alpha\beta} \theta \label{eqn:fvv}
\end{align}
where $\theta(\Delta t,\Delta x)$ is defined by Eq.~\eqref{eqn:fvv}. It approaches $a^2 +1/6$, with $a^2$ defined in Eq.~\eqref{eqn:a2},
which was previously shown in Fig.~10 of~\cite{parsa2017lattice}. What is new here is that we were able to show that this result holds all values of $u$. In the context of the research for this paper we discovered that while the discussion in the paper by Parsa \textit{et al.} focused on the case $u=0$, the dependence on $u$ for different $\theta$ is actually fascinating. While lattice Boltzmann approaches demand that $\theta$ is a constant, this requirement is not consistent with the requirement of a positive distribution $f_i^{eq}$ for small values of $\theta$.  

\begin{figure}
    \centering
    \includegraphics[width=\columnwidth]{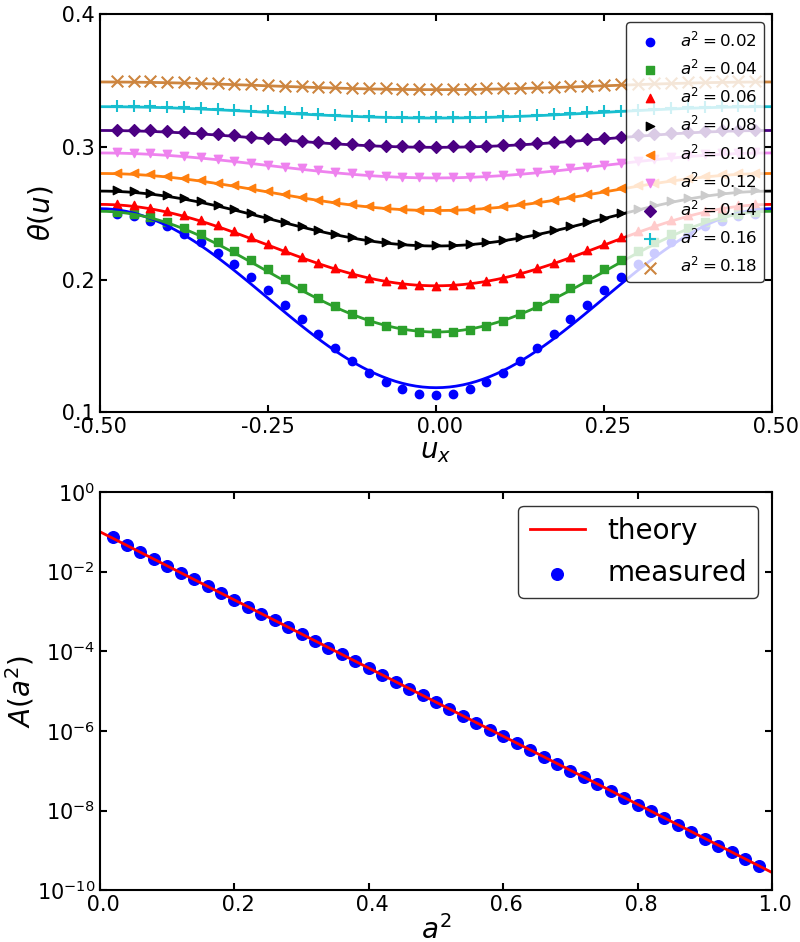}
    \caption{$\theta$ defined in (\ref{eqn:fvv}) as function of mean velocity for different $a^2$ is shown in (a) and the measured amplitude $A(a^2)$ is shown in (b). Both are compared to the numerical fit of Eq.~\eqref{eqn:PjLG}.}
    \label{fig:A}
\end{figure}

Numerical evidence shows that we can approximate 
\begin{equation}
    \theta(u) =\left[ \left(a^2+\frac{1}{6}\right)-A(a^2) \cos(2\pi u_x)\right]
\end{equation}
where
\begin{equation}
    A(a^2) = 0.1  \exp(-19.7 a^2)
    \label{eqn:A}
\end{equation}
where the numerical values were found by numerical fitting, as shown in Fig.~\ref{fig:A}. This shows that while the lattice gas expression for the second moment cannot be strictly Galilean invariant, it exponentially approaches a Galilean invariant value for larger $a^2$.

These results allow us to make an analytical prediction for lattice gas momentum fluctuations for a dilute system. The probability of observing a momentum of $j$ for a lattice gas is given by a combination of Poisson distributed random numbers:
\begin{align}
    P^{LG}(j) =& \sum_{n_1=0}^N\cdots\sum_{n_I=0}^N P(\{n_i\}) \delta\left(j-\sum_i n_i c_i\right).
\end{align}
where $\delta()$ is the delta function and $I$ is the number of discrete lattice velocities.
To evaluate this, it makes sense to first combine all the contributions with a positive velocity $c_{ix}>0$. For this purpose let us define a partial velocity set $c_{ix}^+$ with $V^+$ elements, which contains all velocities with a positive $x$\nobreakdash-component. We then get
\begin{align}
    P(j^+)  =& \sum_{n^+_1=0}^{\widehat{j^+}}\cdots \sum_{n^+_{V^+}=0}^{\widehat{j^+}} \prod_{i=1}^{V^+}  e^{-f_i^{eq}}\frac{(f_i^{eq})^{n_i^+}}{n_i^+ !} \delta(\sum n_i^+ c_{ix}^+-j^+)\nonumber\\
    =&e^{-\sum_{i=1}^{V^+} f_i^+} 
    \frac{(\sum f_i^+)^{\widehat{j^+}}}{\widehat{j^+} !}
\end{align}
where we introduced an integer momentum as $\widehat{j^+}=j^+\Delta t/\Delta x$.
Similarly we get
\begin{align}
    P(j^-)  =& e^{-\sum_{i=1}^{V^-} f_i^-} 
    \frac{(\sum f_i^-)^{\widehat{j^-}}}{\widehat{j^-} !}.
\end{align}
The probability distribution for the full non-dimensional momentum $\widehat{j}=j\Delta t/\Delta x$ is then given by
\begin{align}
    &P^{LG}(\widehat{j})
    \nonumber\\
    =&\sum_{\widehat{j^+}=0}^{\infty}\sum_{\widehat{j^-}=0}^\infty e^{-(f^++f^-)}\frac{(f^+)^{\widehat{j^+}}}{\widehat{j^+}!}\frac{(f^-)^{\widehat{j^-}}}{\widehat{j^-}!} \delta_{(\widehat{j^+}-\widehat{j^-}),\widehat{j}}\nonumber\\
    =&e^{-(f^++f^-)}\sum_{\widehat{j^+}=\max(0,\widehat{j})}^\infty
    \frac{(f^+)^{\widehat{j^+}}}{\widehat{j^+}!} \frac{(f^-)^{\widehat{j^+}-\widehat{j}}}{(\widehat{j^+}-\widehat{j})!}
    \nonumber\\
    =&e^{-(f^++f^-)} \left(\frac{f^+}{f^-}\right)^{\widehat{j}/2}  I_{|\widehat{j}|}(2\sqrt{f^+ f^-})
\end{align}
where $I_j()$ is the modified Bessel function of the first kind. This result will be identical to Eq.~\eqref{eqn:jLGdistbin} when $N_\xi\rightarrow \infty$, i.e. the system becomes infinite. Looking back at Eq.~\eqref{eqn:fv} and Eq.~\eqref{eqn:fvv} we see that we can identify
\begin{align}
    (f^+-f^-)\frac{\Delta x}{\Delta t} &=  \rho u_x\\
    (f^++f^-)\frac{\Delta x^2}{\Delta t^2} &=  \rho u_x^2 +\rho \theta .
\end{align}
So we can write the probability distribution for the current in terms of the equilibrium properties as
\begin{align}
    P^{LG}(\widehat{j}) =& e^{-(\rho u_x^2+\rho\theta)} \left(\frac{u_x+u_x^2+\theta}{-u_x+u_x^2+\theta}\right)^{\widehat{j}/2}  \nonumber\\&
    \times I_{|\widehat{j}|}\sqrt{\rho(u_x^2+\theta)^2-\rho^2 u_x^2}.
    \label{eqn:PjLG}
\end{align}
With this we obtain in MD units
\begin{align}
    \sum_j P^{LG}(j) j &=  \rho u\\
    \sum_j P^{LG}(j) j^2 &= \rho u_x^2 +\rho \theta
\end{align}
as should be expected for consistency with Eqs.~(\ref{eqn:f}--\ref{eqn:fvv}). In particular for the case $u=0$ we predict
\begin{align}
    \langle j^{LG}(\xi,t)^2 \rangle =& \rho^{eq}\theta
    \nonumber\\
    \approx & \rho^{eq} \left(\frac{\langle  \delta x^2\rangle}{\Delta x^2}+1/6\right) \frac{\Delta x^2}{\Delta t^2}
    \nonumber\\
    =&\rho^{eq}\left(\frac{\langle \delta x^2\rangle}{\Delta t^2}+\frac{\Delta x^2}{6\Delta t^2}\right). 
    \label{eqn:jLGdist}
\end{align}
When comparing this to the results for $P^{\Delta t}$ of Eq.~\eqref{eqn:jdt} we see that the first terms are identical, but there is a second term for the Lattice Gas. 
For the case of $a^2=\langle \delta x^2\rangle/\Delta x^2=2/11$ considered in this paper both terms are of equal magnitude, and we predict that the noise amplitude is a little less than twice as large for the lattice gas current than for the current defined by displacements. 

\section{Discussion}
A comparison of the predictions for the current fluctuations with the actually measured current fluctuations for different $\Delta t$ is shown in Fig.~\ref{fig:predict}. The fundamental definition of the momentum $j$ has fluctuations that are independent of $\Delta t$. The time-averaged current $j^{\Delta t}$ fluctuations agree with the fundamental current for small $\Delta t$ where the approximation $v=\delta x/\Delta t$ holds, but then starts to be reduced. The actually measured fluctuations of this current, however, are less reduced than our simple theory predicts. This indicates that particles displacements fail to be independent of each other as particles collide with each other. This effect appears to be very strong, and as soon as a deviation from the ballistic regime is seen, the prediction based on independent displacements fails to be accurate. 

Now the fluctuations for the lattice gas current $j^{LG}$ are more puzzling, as they fail to converge with the fundamental current fluctuations, even for small $\Delta t$. They do, however, agree well with the experimentally measured lattice  gas current fluctuations. For large $\Delta t$ the measured displacement and lattice gas currents start to converge. 

\begin{figure}
    \centering
    \includegraphics[width=\columnwidth]{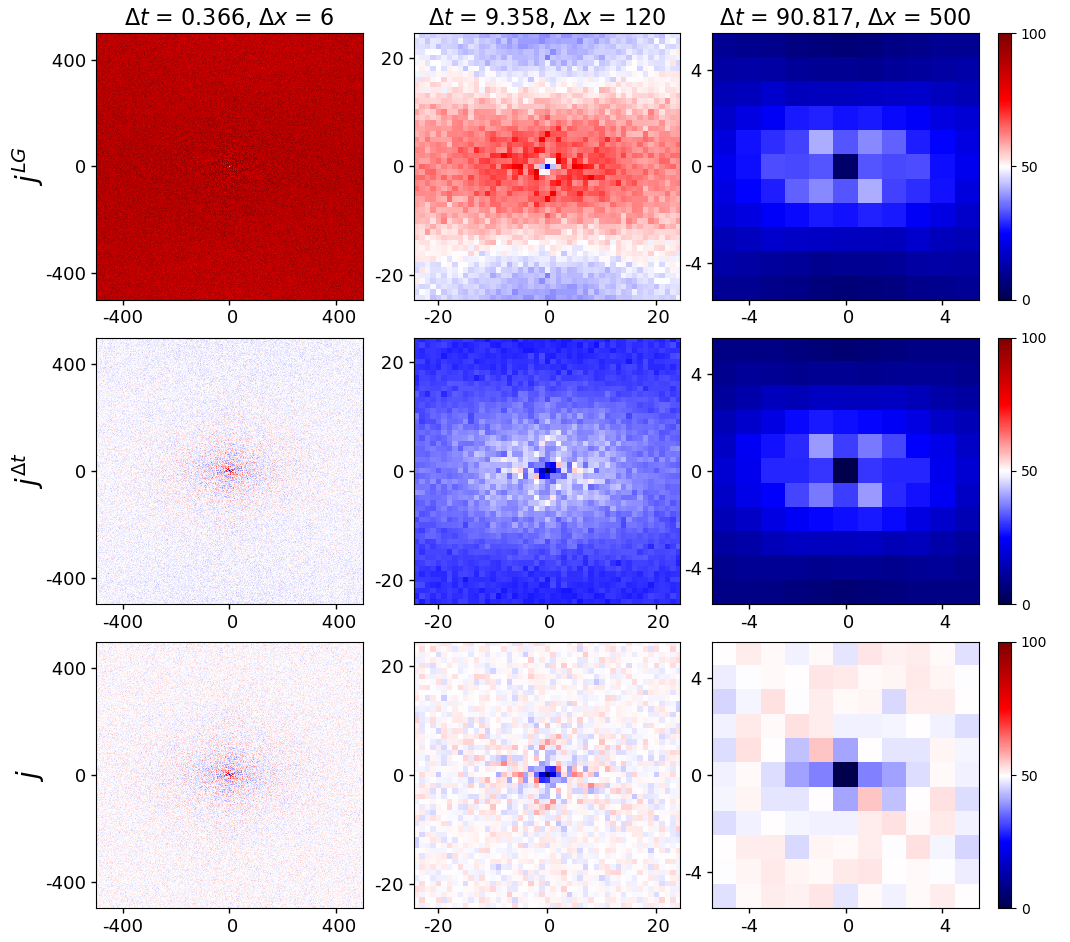}
    \caption{Fourier transforms $\langle j_x^a(k) j_x^a(-k)$ for three different current definitions $j^a=\{j,j^{\Delta t},j^{LG}\}$ considered in this paper for three different time-steps $\Delta t$.}
    \label{fig:fourier}
\end{figure}

We found these results very puzzling, since we expected that currents should agree in the limit of small times. However, our next possible explanation was that the existence of a lattice can introduce effects at small wavelength, whereas we would expect long wave\nobreakdash-length fluctuations, which evolve much slower, should be in agreement. This would also indicate that in the hydrodynamic limit all currents should agree. We therefore looked at the Fourier transforms of the current fluctuations for all three currents at three different levels of coarse graining. The results are shown in Fig.~\ref{fig:fourier}. As expected for the dilute system under consideration, the fluctuations for the $x$\nobreakdash-component of the current $j$ are always flat. The Fourier plot for the time-averaged current $j^{\Delta t}_x$ is identical for small $\Delta t$, but for larger $\Delta t$ the current diminishes for fast modes corresponding to short wavelength, and this effect is clearly visible in the graph. It is also obvious that current in the orthogonal $y$\nobreakdash-direction decays more quickly whereas the value for slowly evolving large wavelength (small $k$) remains essentially unaffected by the time-averaging.

Now the Fourier transform of the lattice gas current is also $k$\nobreakdash-independent, but at the much higher fluctuation amplitudes as predicted by Eq.~\eqref{eqn:jLGdist}. This implies that the lattice gas discretization changes the fluctuation amplitude of the noise-component of the current even for very large wavelength, while not affecting the ensemble averaged current. This result was completely unexpected, since we anticipated that all three momentum measures would agree in the thermodynamic limit. Surprisingly this is obviously not the case.

For the larger time-step of $\Delta t=9.358$ we saw in Fig.~\ref{fig:predict} that the total fluctuation amplitude still tracks the analytically predicted one very closely. Here we see, however, that the Fourier spectrum is no longer flat but instead decays, particularly in the orthogonal direction. At the even larger time-step of $\Delta t=90.817$ we saw in Fig.~\ref{fig:predict} that the analytical prediction and the measured fluctuations no longer agree. However, we also saw that the overall fluctuations of $j^{\Delta t}$ and $j^{LG}$ converged. Here we now observe that even the full Fourier representations of the two measures agree. 

In conclusion, we have discussed three definitions of momentum to examine the effect of different coarse-graining procedures on the definition of the fundamental conserved quantity of momentum. Surprisingly we found that different coarse-grained definitions of the momentum can disagree for the fluctuating components, even in the hydrodynamic limit, while ensemble averages of the momenta fully agree. This strange result is in full agreement with our analytical predictions, which is evidence that this strange result is not spurious. We saw that for short coarse-graining times $\Delta t$ the lattice gas does obey the predictions for fluctuations underlying the work of Adhikari \textit{et al.}~\cite{adhikari2005fluctuating} as well as later research~\cite{dunweg2007statistical,kaehler2013fluctuating,wagner2016fluctuating,blommel2018integer}, but the momentum fluctuations are about twice as large as the fluctuations of the fundamental definition of momentum of a lattice cell.

Another result that requires further investigation is that the lattice gas momentum starts to converge towards another fundamental time-averaged definition of momentum for larger $\Delta t$, and in this limit numerical evidence suggests that the momentum fluctuations for large wavelength now agree with the fluctuations corresponding the fundamental definition of momentum.

\bibliography{paper_momentum.bib}

\begin{thebibliography}{36}%
\makeatletter
\providecommand \@ifxundefined [1]{%
 \@ifx{#1\undefined}
}%
\providecommand \@ifnum [1]{%
 \ifnum #1\expandafter \@firstoftwo
 \else \expandafter \@secondoftwo
 \fi
}%
\providecommand \@ifx [1]{%
 \ifx #1\expandafter \@firstoftwo
 \else \expandafter \@secondoftwo
 \fi
}%
\providecommand \natexlab [1]{#1}%
\providecommand \enquote  [1]{``#1''}%
\providecommand \bibnamefont  [1]{#1}%
\providecommand \bibfnamefont [1]{#1}%
\providecommand \citenamefont [1]{#1}%
\providecommand \href@noop [0]{\@secondoftwo}%
\providecommand \href [0]{\begingroup \@sanitize@url \@href}%
\providecommand \@href[1]{\@@startlink{#1}\@@href}%
\providecommand \@@href[1]{\endgroup#1\@@endlink}%
\providecommand \@sanitize@url [0]{\catcode `\\12\catcode `\$12\catcode
  `\&12\catcode `\#12\catcode `\^12\catcode `\_12\catcode `\%12\relax}%
\providecommand \@@startlink[1]{}%
\providecommand \@@endlink[0]{}%
\providecommand \url  [0]{\begingroup\@sanitize@url \@url }%
\providecommand \@url [1]{\endgroup\@href {#1}{\urlprefix }}%
\providecommand \urlprefix  [0]{URL }%
\providecommand \Eprint [0]{\href }%
\providecommand \doibase [0]{https://doi.org/}%
\providecommand \selectlanguage [0]{\@gobble}%
\providecommand \bibinfo  [0]{\@secondoftwo}%
\providecommand \bibfield  [0]{\@secondoftwo}%
\providecommand \translation [1]{[#1]}%
\providecommand \BibitemOpen [0]{}%
\providecommand \bibitemStop [0]{}%
\providecommand \bibitemNoStop [0]{.\EOS\space}%
\providecommand \EOS [0]{\spacefactor3000\relax}%
\providecommand \BibitemShut  [1]{\csname bibitem#1\endcsname}%
\let\auto@bib@innerbib\@empty
\bibitem [{\citenamefont {Gorban}(2006)}]{Gorban2006}%
  \BibitemOpen
  \bibfield  {author} {\bibinfo {author} {\bibfnamefont {A.~N.}\ \bibnamefont
  {Gorban}},\ }\bibinfo {title} {Basic types of coarse-graining},\ in\ \href
  {https://doi.org/10.1007/3-540-35888-9_7} {\emph {\bibinfo {booktitle} {Model
  Reduction and Coarse-Graining Approaches for Multiscale Phenomena}}},\
  \bibinfo {editor} {edited by\ \bibinfo {editor} {\bibfnamefont {A.~N.}\
  \bibnamefont {Gorban}}, \bibinfo {editor} {\bibfnamefont {I.~G.}\
  \bibnamefont {Kevrekidis}}, \bibinfo {editor} {\bibfnamefont
  {C.}~\bibnamefont {Theodoropoulos}}, \bibinfo {editor} {\bibfnamefont
  {N.~K.}\ \bibnamefont {Kazantzis}},\ and\ \bibinfo {editor} {\bibfnamefont
  {H.~C.}\ \bibnamefont {{\"O}ttinger}}}\ (\bibinfo  {publisher} {Springer
  Berlin Heidelberg},\ \bibinfo {address} {Berlin, Heidelberg},\ \bibinfo
  {year} {2006})\ pp.\ \bibinfo {pages} {117--176}\BibitemShut {NoStop}%
\bibitem [{\citenamefont {Parsa}\ and\ \citenamefont
  {Wagner}(2020)}]{parsa2020large}%
  \BibitemOpen
  \bibfield  {author} {\bibinfo {author} {\bibfnamefont {M.~R.}\ \bibnamefont
  {Parsa}}\ and\ \bibinfo {author} {\bibfnamefont {A.~J.}\ \bibnamefont
  {Wagner}},\ }\bibfield  {title} {\bibinfo {title} {Large fluctuations in
  nonideal coarse-grained systems},\ }\href@noop {} {\bibfield  {journal}
  {\bibinfo  {journal} {Physical Review Letters}\ }\textbf {\bibinfo {volume}
  {124}},\ \bibinfo {pages} {234501} (\bibinfo {year} {2020})}\BibitemShut
  {NoStop}%
\bibitem [{\citenamefont {Parsa}\ and\ \citenamefont
  {Wagner}(2017)}]{parsa2017lattice}%
  \BibitemOpen
  \bibfield  {author} {\bibinfo {author} {\bibfnamefont {M.~R.}\ \bibnamefont
  {Parsa}}\ and\ \bibinfo {author} {\bibfnamefont {A.~J.}\ \bibnamefont
  {Wagner}},\ }\bibfield  {title} {\bibinfo {title} {Lattice gas with molecular
  dynamics collision operator},\ }\href@noop {} {\bibfield  {journal} {\bibinfo
   {journal} {Physical Review E}\ }\textbf {\bibinfo {volume} {96}},\ \bibinfo
  {pages} {013314} (\bibinfo {year} {2017})}\BibitemShut {NoStop}%
\bibitem [{\citenamefont {Adhikari}\ \emph {et~al.}(2005)\citenamefont
  {Adhikari}, \citenamefont {Stratford}, \citenamefont {Cates},\ and\
  \citenamefont {Wagner}}]{adhikari2005fluctuating}%
  \BibitemOpen
  \bibfield  {author} {\bibinfo {author} {\bibfnamefont {R.}~\bibnamefont
  {Adhikari}}, \bibinfo {author} {\bibfnamefont {K.}~\bibnamefont {Stratford}},
  \bibinfo {author} {\bibfnamefont {M.}~\bibnamefont {Cates}},\ and\ \bibinfo
  {author} {\bibfnamefont {A.}~\bibnamefont {Wagner}},\ }\bibfield  {title}
  {\bibinfo {title} {Fluctuating lattice boltzmann},\ }\href@noop {} {\bibfield
   {journal} {\bibinfo  {journal} {EPL (Europhysics Letters)}\ }\textbf
  {\bibinfo {volume} {71}},\ \bibinfo {pages} {473} (\bibinfo {year}
  {2005})}\BibitemShut {NoStop}%
\bibitem [{\citenamefont {Frisch}\ \emph {et~al.}(1986)\citenamefont {Frisch},
  \citenamefont {Hasslacher},\ and\ \citenamefont
  {Pomeau}}]{frisch1986lattice}%
  \BibitemOpen
  \bibfield  {author} {\bibinfo {author} {\bibfnamefont {U.}~\bibnamefont
  {Frisch}}, \bibinfo {author} {\bibfnamefont {B.}~\bibnamefont {Hasslacher}},\
  and\ \bibinfo {author} {\bibfnamefont {Y.}~\bibnamefont {Pomeau}},\
  }\bibfield  {title} {\bibinfo {title} {Lattice-gas automata for the
  navier-stokes equation},\ }\href@noop {} {\bibfield  {journal} {\bibinfo
  {journal} {Physical Review Letters}\ }\textbf {\bibinfo {volume} {56}},\
  \bibinfo {pages} {1505} (\bibinfo {year} {1986})}\BibitemShut {NoStop}%
\bibitem [{\citenamefont {Boon}\ and\ \citenamefont
  {Yip}(1991)}]{boon1991molecular}%
  \BibitemOpen
  \bibfield  {author} {\bibinfo {author} {\bibfnamefont {J.~P.}\ \bibnamefont
  {Boon}}\ and\ \bibinfo {author} {\bibfnamefont {S.}~\bibnamefont {Yip}},\
  }\href@noop {} {\emph {\bibinfo {title} {Molecular hydrodynamics}}}\
  (\bibinfo  {publisher} {Courier Corporation},\ \bibinfo {year}
  {1991})\BibitemShut {NoStop}%
\bibitem [{\citenamefont {Ladd}(1993)}]{ladd1993short}%
  \BibitemOpen
  \bibfield  {author} {\bibinfo {author} {\bibfnamefont {A.~J.}\ \bibnamefont
  {Ladd}},\ }\bibfield  {title} {\bibinfo {title} {Short-time motion of
  colloidal particles: Numerical simulation via a fluctuating lattice-boltzmann
  equation},\ }\href@noop {} {\bibfield  {journal} {\bibinfo  {journal}
  {Physical Review Letters}\ }\textbf {\bibinfo {volume} {70}},\ \bibinfo
  {pages} {1339} (\bibinfo {year} {1993})}\BibitemShut {NoStop}%
\bibitem [{\citenamefont {D{\"u}nweg}\ \emph {et~al.}(2007)\citenamefont
  {D{\"u}nweg}, \citenamefont {Schiller},\ and\ \citenamefont
  {Ladd}}]{dunweg2007statistical}%
  \BibitemOpen
  \bibfield  {author} {\bibinfo {author} {\bibfnamefont {B.}~\bibnamefont
  {D{\"u}nweg}}, \bibinfo {author} {\bibfnamefont {U.~D.}\ \bibnamefont
  {Schiller}},\ and\ \bibinfo {author} {\bibfnamefont {A.~J.}\ \bibnamefont
  {Ladd}},\ }\bibfield  {title} {\bibinfo {title} {Statistical mechanics of the
  fluctuating lattice boltzmann equation},\ }\href@noop {} {\bibfield
  {journal} {\bibinfo  {journal} {Physical Review E}\ }\textbf {\bibinfo
  {volume} {76}},\ \bibinfo {pages} {036704} (\bibinfo {year}
  {2007})}\BibitemShut {NoStop}%
\bibitem [{\citenamefont {Grosfils}\ \emph {et~al.}(1992)\citenamefont
  {Grosfils}, \citenamefont {Boon},\ and\ \citenamefont
  {Lallemand}}]{grosfils1992spontaneous}%
  \BibitemOpen
  \bibfield  {author} {\bibinfo {author} {\bibfnamefont {P.}~\bibnamefont
  {Grosfils}}, \bibinfo {author} {\bibfnamefont {J.-P.}\ \bibnamefont {Boon}},\
  and\ \bibinfo {author} {\bibfnamefont {P.}~\bibnamefont {Lallemand}},\
  }\bibfield  {title} {\bibinfo {title} {Spontaneous fluctuation correlations
  in thermal lattice-gas automata},\ }\href@noop {} {\bibfield  {journal}
  {\bibinfo  {journal} {Physical Review Letters}\ }\textbf {\bibinfo {volume}
  {68}},\ \bibinfo {pages} {1077} (\bibinfo {year} {1992})}\BibitemShut
  {NoStop}%
\bibitem [{\citenamefont {Hoogerbrugge}\ and\ \citenamefont
  {Koelman}(1992)}]{hoogerbrugge1992simulating}%
  \BibitemOpen
  \bibfield  {author} {\bibinfo {author} {\bibfnamefont {P.}~\bibnamefont
  {Hoogerbrugge}}\ and\ \bibinfo {author} {\bibfnamefont {J.}~\bibnamefont
  {Koelman}},\ }\bibfield  {title} {\bibinfo {title} {Simulating microscopic
  hydrodynamic phenomena with dissipative particle dynamics},\ }\href@noop {}
  {\bibfield  {journal} {\bibinfo  {journal} {EPL (Europhysics Letters)}\
  }\textbf {\bibinfo {volume} {19}},\ \bibinfo {pages} {155} (\bibinfo {year}
  {1992})}\BibitemShut {NoStop}%
\bibitem [{\citenamefont {Espanol}\ and\ \citenamefont
  {Warren}(1995)}]{espanol1995statistical}%
  \BibitemOpen
  \bibfield  {author} {\bibinfo {author} {\bibfnamefont {P.}~\bibnamefont
  {Espanol}}\ and\ \bibinfo {author} {\bibfnamefont {P.}~\bibnamefont
  {Warren}},\ }\bibfield  {title} {\bibinfo {title} {Statistical mechanics of
  dissipative particle dynamics},\ }\href@noop {} {\bibfield  {journal}
  {\bibinfo  {journal} {EPL (Europhysics Letters)}\ }\textbf {\bibinfo {volume}
  {30}},\ \bibinfo {pages} {191} (\bibinfo {year} {1995})}\BibitemShut
  {NoStop}%
\bibitem [{\citenamefont {Ihle}\ and\ \citenamefont
  {Kroll}(2001)}]{ihle2001stochastic}%
  \BibitemOpen
  \bibfield  {author} {\bibinfo {author} {\bibfnamefont {T.}~\bibnamefont
  {Ihle}}\ and\ \bibinfo {author} {\bibfnamefont {D.}~\bibnamefont {Kroll}},\
  }\bibfield  {title} {\bibinfo {title} {Stochastic rotation dynamics: A
  galilean-invariant mesoscopic model for fluid flow},\ }\href@noop {}
  {\bibfield  {journal} {\bibinfo  {journal} {Physical Review E}\ }\textbf
  {\bibinfo {volume} {63}},\ \bibinfo {pages} {020201} (\bibinfo {year}
  {2001})}\BibitemShut {NoStop}%
\bibitem [{\citenamefont {Ihle}\ and\ \citenamefont
  {Kroll}(2003)}]{ihle2003stochastic}%
  \BibitemOpen
  \bibfield  {author} {\bibinfo {author} {\bibfnamefont {T.}~\bibnamefont
  {Ihle}}\ and\ \bibinfo {author} {\bibfnamefont {D.~M.}\ \bibnamefont
  {Kroll}},\ }\bibfield  {title} {\bibinfo {title} {Stochastic rotation
  dynamics. i. formalism, galilean invariance, and green-kubo relations},\
  }\href@noop {} {\bibfield  {journal} {\bibinfo  {journal} {Physical Review
  E}\ }\textbf {\bibinfo {volume} {67}},\ \bibinfo {pages} {066705} (\bibinfo
  {year} {2003})}\BibitemShut {NoStop}%
\bibitem [{\citenamefont {T{\"u}zel}\ \emph {et~al.}(2003)\citenamefont
  {T{\"u}zel}, \citenamefont {Strauss}, \citenamefont {Ihle},\ and\
  \citenamefont {Kroll}}]{tuzel2003transport}%
  \BibitemOpen
  \bibfield  {author} {\bibinfo {author} {\bibfnamefont {E.}~\bibnamefont
  {T{\"u}zel}}, \bibinfo {author} {\bibfnamefont {M.}~\bibnamefont {Strauss}},
  \bibinfo {author} {\bibfnamefont {T.}~\bibnamefont {Ihle}},\ and\ \bibinfo
  {author} {\bibfnamefont {D.~M.}\ \bibnamefont {Kroll}},\ }\bibfield  {title}
  {\bibinfo {title} {Transport coefficients for stochastic rotation dynamics in
  three dimensions},\ }\href@noop {} {\bibfield  {journal} {\bibinfo  {journal}
  {Physical Review E}\ }\textbf {\bibinfo {volume} {68}},\ \bibinfo {pages}
  {036701} (\bibinfo {year} {2003})}\BibitemShut {NoStop}%
\bibitem [{\citenamefont {Wolfram}(1986)}]{wolfram1986cellular}%
  \BibitemOpen
  \bibfield  {author} {\bibinfo {author} {\bibfnamefont {S.}~\bibnamefont
  {Wolfram}},\ }\bibfield  {title} {\bibinfo {title} {Cellular automaton fluids
  1: Basic theory},\ }\href@noop {} {\bibfield  {journal} {\bibinfo  {journal}
  {Journal of statistical physics}\ }\textbf {\bibinfo {volume} {45}},\
  \bibinfo {pages} {471} (\bibinfo {year} {1986})}\BibitemShut {NoStop}%
\bibitem [{\citenamefont {Ladd}\ \emph {et~al.}(1988)\citenamefont {Ladd},
  \citenamefont {Colvin},\ and\ \citenamefont {Frenkel}}]{ladd1988application}%
  \BibitemOpen
  \bibfield  {author} {\bibinfo {author} {\bibfnamefont {A.~J.}\ \bibnamefont
  {Ladd}}, \bibinfo {author} {\bibfnamefont {M.~E.}\ \bibnamefont {Colvin}},\
  and\ \bibinfo {author} {\bibfnamefont {D.}~\bibnamefont {Frenkel}},\
  }\bibfield  {title} {\bibinfo {title} {Application of lattice-gas cellular
  automata to the brownian motion of solids in suspension},\ }\href@noop {}
  {\bibfield  {journal} {\bibinfo  {journal} {Physical review letters}\
  }\textbf {\bibinfo {volume} {60}},\ \bibinfo {pages} {975} (\bibinfo {year}
  {1988})}\BibitemShut {NoStop}%
\bibitem [{\citenamefont {Ladd}\ and\ \citenamefont
  {Frenkel}(1990)}]{ladd1990dissipative}%
  \BibitemOpen
  \bibfield  {author} {\bibinfo {author} {\bibfnamefont {A.~J.}\ \bibnamefont
  {Ladd}}\ and\ \bibinfo {author} {\bibfnamefont {D.}~\bibnamefont {Frenkel}},\
  }\bibfield  {title} {\bibinfo {title} {Dissipative hydrodynamic interactions
  via lattice-gas cellular automata},\ }\href@noop {} {\bibfield  {journal}
  {\bibinfo  {journal} {Physics of fluids A: fluid dynamics}\ }\textbf
  {\bibinfo {volume} {2}},\ \bibinfo {pages} {1921} (\bibinfo {year}
  {1990})}\BibitemShut {NoStop}%
\bibitem [{\citenamefont {Rivet}\ and\ \citenamefont
  {Boon}(2005)}]{rivet2005lattice}%
  \BibitemOpen
  \bibfield  {author} {\bibinfo {author} {\bibfnamefont {J.-P.}\ \bibnamefont
  {Rivet}}\ and\ \bibinfo {author} {\bibfnamefont {J.-P.}\ \bibnamefont
  {Boon}},\ }\href@noop {} {\emph {\bibinfo {title} {Lattice gas
  hydrodynamics}}},\ Vol.~\bibinfo {volume} {11}\ (\bibinfo  {publisher}
  {Cambridge University Press},\ \bibinfo {year} {2005})\BibitemShut {NoStop}%
\bibitem [{\citenamefont {Frenkel}\ and\ \citenamefont
  {Ernst}(1989)}]{frenkel1989simulation}%
  \BibitemOpen
  \bibfield  {author} {\bibinfo {author} {\bibfnamefont {D.}~\bibnamefont
  {Frenkel}}\ and\ \bibinfo {author} {\bibfnamefont {M.}~\bibnamefont
  {Ernst}},\ }\bibfield  {title} {\bibinfo {title} {Simulation of diffusion in
  a two-dimensional lattice-gas cellular automaton: a test of mode-coupling
  theory},\ }\href@noop {} {\bibfield  {journal} {\bibinfo  {journal} {Physical
  review letters}\ }\textbf {\bibinfo {volume} {63}},\ \bibinfo {pages} {2165}
  (\bibinfo {year} {1989})}\BibitemShut {NoStop}%
\bibitem [{\citenamefont {Higuera}\ and\ \citenamefont
  {Jim{\'e}nez}(1989)}]{higuera1989boltzmann}%
  \BibitemOpen
  \bibfield  {author} {\bibinfo {author} {\bibfnamefont {F.~J.}\ \bibnamefont
  {Higuera}}\ and\ \bibinfo {author} {\bibfnamefont {J.}~\bibnamefont
  {Jim{\'e}nez}},\ }\bibfield  {title} {\bibinfo {title} {Boltzmann approach to
  lattice gas simulations},\ }\href@noop {} {\bibfield  {journal} {\bibinfo
  {journal} {EPL (Europhysics Letters)}\ }\textbf {\bibinfo {volume} {9}},\
  \bibinfo {pages} {663} (\bibinfo {year} {1989})}\BibitemShut {NoStop}%
\bibitem [{\citenamefont {Higuera}\ \emph {et~al.}(1989)\citenamefont
  {Higuera}, \citenamefont {Succi},\ and\ \citenamefont
  {Benzi}}]{higuera1989lattice}%
  \BibitemOpen
  \bibfield  {author} {\bibinfo {author} {\bibfnamefont {F.}~\bibnamefont
  {Higuera}}, \bibinfo {author} {\bibfnamefont {S.}~\bibnamefont {Succi}},\
  and\ \bibinfo {author} {\bibfnamefont {R.}~\bibnamefont {Benzi}},\ }\bibfield
   {title} {\bibinfo {title} {Lattice gas dynamics with enhanced collisions},\
  }\href@noop {} {\bibfield  {journal} {\bibinfo  {journal} {EPL (Europhysics
  Letters)}\ }\textbf {\bibinfo {volume} {9}},\ \bibinfo {pages} {345}
  (\bibinfo {year} {1989})}\BibitemShut {NoStop}%
\bibitem [{\citenamefont {Qian}\ \emph {et~al.}(1992)\citenamefont {Qian},
  \citenamefont {d'Humi{\`e}res},\ and\ \citenamefont
  {Lallemand}}]{qian1992lattice}%
  \BibitemOpen
  \bibfield  {author} {\bibinfo {author} {\bibfnamefont {Y.-H.}\ \bibnamefont
  {Qian}}, \bibinfo {author} {\bibfnamefont {D.}~\bibnamefont
  {d'Humi{\`e}res}},\ and\ \bibinfo {author} {\bibfnamefont {P.}~\bibnamefont
  {Lallemand}},\ }\bibfield  {title} {\bibinfo {title} {Lattice bgk models for
  navier-stokes equation},\ }\href@noop {} {\bibfield  {journal} {\bibinfo
  {journal} {EPL (Europhysics Letters)}\ }\textbf {\bibinfo {volume} {17}},\
  \bibinfo {pages} {479} (\bibinfo {year} {1992})}\BibitemShut {NoStop}%
\bibitem [{\citenamefont {Ladd}(1994)}]{ladd1994numerical}%
  \BibitemOpen
  \bibfield  {author} {\bibinfo {author} {\bibfnamefont {A.~J.}\ \bibnamefont
  {Ladd}},\ }\bibfield  {title} {\bibinfo {title} {Numerical simulations of
  particulate suspensions via a discretized boltzmann equation. part 1.
  theoretical foundation},\ }\href@noop {} {\bibfield  {journal} {\bibinfo
  {journal} {Journal of fluid mechanics}\ }\textbf {\bibinfo {volume} {271}},\
  \bibinfo {pages} {285} (\bibinfo {year} {1994})}\BibitemShut {NoStop}%
\bibitem [{\citenamefont {Kaehler}\ and\ \citenamefont
  {Wagner}(2013)}]{kaehler2013fluctuating}%
  \BibitemOpen
  \bibfield  {author} {\bibinfo {author} {\bibfnamefont {G.}~\bibnamefont
  {Kaehler}}\ and\ \bibinfo {author} {\bibfnamefont {A.}~\bibnamefont
  {Wagner}},\ }\bibfield  {title} {\bibinfo {title} {Fluctuating ideal-gas
  lattice boltzmann method with fluctuation dissipation theorem for
  nonvanishing velocities},\ }\href@noop {} {\bibfield  {journal} {\bibinfo
  {journal} {Physical Review E}\ }\textbf {\bibinfo {volume} {87}},\ \bibinfo
  {pages} {063310} (\bibinfo {year} {2013})}\BibitemShut {NoStop}%
\bibitem [{\citenamefont {Blommel}\ and\ \citenamefont
  {Wagner}(2018)}]{blommel2018integer}%
  \BibitemOpen
  \bibfield  {author} {\bibinfo {author} {\bibfnamefont {T.}~\bibnamefont
  {Blommel}}\ and\ \bibinfo {author} {\bibfnamefont {A.~J.}\ \bibnamefont
  {Wagner}},\ }\bibfield  {title} {\bibinfo {title} {Integer lattice gas with
  monte carlo collision operator recovers the lattice boltzmann method with
  poisson-distributed fluctuations},\ }\href@noop {} {\bibfield  {journal}
  {\bibinfo  {journal} {Physical Review E}\ }\textbf {\bibinfo {volume} {97}},\
  \bibinfo {pages} {023310} (\bibinfo {year} {2018})}\BibitemShut {NoStop}%
\bibitem [{\citenamefont {Ansumali}\ \emph {et~al.}(2003)\citenamefont
  {Ansumali}, \citenamefont {Karlin},\ and\ \citenamefont
  {{\"O}ttinger}}]{ansumali2003minimal}%
  \BibitemOpen
  \bibfield  {author} {\bibinfo {author} {\bibfnamefont {S.}~\bibnamefont
  {Ansumali}}, \bibinfo {author} {\bibfnamefont {I.~V.}\ \bibnamefont
  {Karlin}},\ and\ \bibinfo {author} {\bibfnamefont {H.~C.}\ \bibnamefont
  {{\"O}ttinger}},\ }\bibfield  {title} {\bibinfo {title} {Minimal entropic
  kinetic models for hydrodynamics},\ }\href@noop {} {\bibfield  {journal}
  {\bibinfo  {journal} {EPL (Europhysics Letters)}\ }\textbf {\bibinfo {volume}
  {63}},\ \bibinfo {pages} {798} (\bibinfo {year} {2003})}\BibitemShut
  {NoStop}%
\bibitem [{\citenamefont {Espa{\~n}ol}\ \emph {et~al.}(1997)\citenamefont
  {Espa{\~n}ol}, \citenamefont {Serrano},\ and\ \citenamefont
  {Zu{\~n}iga}}]{espanol1997coarse}%
  \BibitemOpen
  \bibfield  {author} {\bibinfo {author} {\bibfnamefont {P.}~\bibnamefont
  {Espa{\~n}ol}}, \bibinfo {author} {\bibfnamefont {M.}~\bibnamefont
  {Serrano}},\ and\ \bibinfo {author} {\bibfnamefont {I.}~\bibnamefont
  {Zu{\~n}iga}},\ }\bibfield  {title} {\bibinfo {title} {Coarse-graining of a
  fluid and its relation with dissipative particle dynamics and smoothed
  particle dynamic},\ }\href@noop {} {\bibfield  {journal} {\bibinfo  {journal}
  {International Journal of Modern Physics C}\ }\textbf {\bibinfo {volume}
  {8}},\ \bibinfo {pages} {899} (\bibinfo {year} {1997})}\BibitemShut {NoStop}%
\bibitem [{\citenamefont {Rudd}\ and\ \citenamefont
  {Broughton}(1998)}]{rudd1998coarse}%
  \BibitemOpen
  \bibfield  {author} {\bibinfo {author} {\bibfnamefont {R.~E.}\ \bibnamefont
  {Rudd}}\ and\ \bibinfo {author} {\bibfnamefont {J.~Q.}\ \bibnamefont
  {Broughton}},\ }\bibfield  {title} {\bibinfo {title} {Coarse-grained
  molecular dynamics and the atomic limit of finite elements},\ }\href@noop {}
  {\bibfield  {journal} {\bibinfo  {journal} {Physical Review B}\ }\textbf
  {\bibinfo {volume} {58}},\ \bibinfo {pages} {R5893} (\bibinfo {year}
  {1998})}\BibitemShut {NoStop}%
\bibitem [{\citenamefont {Plimpton}(1995)}]{plimpton1995fast}%
  \BibitemOpen
  \bibfield  {author} {\bibinfo {author} {\bibfnamefont {S.}~\bibnamefont
  {Plimpton}},\ }\bibfield  {title} {\bibinfo {title} {Fast parallel algorithms
  for short-range molecular dynamics},\ }\href@noop {} {\bibfield  {journal}
  {\bibinfo  {journal} {Journal of computational physics}\ }\textbf {\bibinfo
  {volume} {117}},\ \bibinfo {pages} {1} (\bibinfo {year} {1995})}\BibitemShut
  {NoStop}%
\bibitem [{\citenamefont {Godoy}\ \emph {et~al.}(2020)\citenamefont {Godoy},
  \citenamefont {Podhorszki}, \citenamefont {Wang}, \citenamefont {Atkins},
  \citenamefont {Eisenhauer}, \citenamefont {Gu}, \citenamefont {Davis},
  \citenamefont {Choi}, \citenamefont {Germaschewski}, \citenamefont {Huck}
  \emph {et~al.}}]{godoy2020adios}%
  \BibitemOpen
  \bibfield  {author} {\bibinfo {author} {\bibfnamefont {W.~F.}\ \bibnamefont
  {Godoy}}, \bibinfo {author} {\bibfnamefont {N.}~\bibnamefont {Podhorszki}},
  \bibinfo {author} {\bibfnamefont {R.}~\bibnamefont {Wang}}, \bibinfo {author}
  {\bibfnamefont {C.}~\bibnamefont {Atkins}}, \bibinfo {author} {\bibfnamefont
  {G.}~\bibnamefont {Eisenhauer}}, \bibinfo {author} {\bibfnamefont
  {J.}~\bibnamefont {Gu}}, \bibinfo {author} {\bibfnamefont {P.}~\bibnamefont
  {Davis}}, \bibinfo {author} {\bibfnamefont {J.}~\bibnamefont {Choi}},
  \bibinfo {author} {\bibfnamefont {K.}~\bibnamefont {Germaschewski}}, \bibinfo
  {author} {\bibfnamefont {K.}~\bibnamefont {Huck}}, \emph {et~al.},\
  }\bibfield  {title} {\bibinfo {title} {Adios 2: The adaptable input output
  system. a framework for high-performance data management},\ }\href@noop {}
  {\bibfield  {journal} {\bibinfo  {journal} {SoftwareX}\ }\textbf {\bibinfo
  {volume} {12}},\ \bibinfo {pages} {100561} (\bibinfo {year}
  {2020})}\BibitemShut {NoStop}%
\bibitem [{\citenamefont {Pachalieva}\ and\ \citenamefont
  {Wagner}(2020)}]{pachalieva2020non}%
  \BibitemOpen
  \bibfield  {author} {\bibinfo {author} {\bibfnamefont {A.}~\bibnamefont
  {Pachalieva}}\ and\ \bibinfo {author} {\bibfnamefont {A.~J.}\ \bibnamefont
  {Wagner}},\ }\bibfield  {title} {\bibinfo {title} {Non-gaussian distribution
  of displacements for lennard-jones particles in equilibrium},\ }\href@noop {}
  {\bibfield  {journal} {\bibinfo  {journal} {arXiv preprint arXiv:2006.05517}\
  } (\bibinfo {year} {2020})}\BibitemShut {NoStop}%
\bibitem [{\citenamefont {Lifschitz}\ and\ \citenamefont
  {Pitajewski}(1983)}]{lifschitz1983physical}%
  \BibitemOpen
  \bibfield  {author} {\bibinfo {author} {\bibfnamefont {E.}~\bibnamefont
  {Lifschitz}}\ and\ \bibinfo {author} {\bibfnamefont {L.}~\bibnamefont
  {Pitajewski}},\ }\bibfield  {title} {\bibinfo {title} {Physical kinetics},\
  }in\ \href@noop {} {\emph {\bibinfo {booktitle} {Textbook of theoretical
  physics. 10}}}\ (\bibinfo {year} {1983})\BibitemShut {NoStop}%
\bibitem [{\citenamefont {Parsa}\ \emph {et~al.}(2019)\citenamefont {Parsa},
  \citenamefont {Pachalieva},\ and\ \citenamefont
  {Wagner}}]{parsa2019validity}%
  \BibitemOpen
  \bibfield  {author} {\bibinfo {author} {\bibfnamefont {M.~R.}\ \bibnamefont
  {Parsa}}, \bibinfo {author} {\bibfnamefont {A.}~\bibnamefont {Pachalieva}},\
  and\ \bibinfo {author} {\bibfnamefont {A.~J.}\ \bibnamefont {Wagner}},\
  }\bibfield  {title} {\bibinfo {title} {Validity of the
  molecular-dynamics-lattice-gas global equilibrium distribution function},\
  }\href@noop {} {\bibfield  {journal} {\bibinfo  {journal} {International
  Journal of Modern Physics C}\ }\textbf {\bibinfo {volume} {30}},\ \bibinfo
  {pages} {1941007} (\bibinfo {year} {2019})}\BibitemShut {NoStop}%
\bibitem [{\citenamefont {Parsa}(2018)}]{parsa2018lattice}%
  \BibitemOpen
  \bibfield  {author} {\bibinfo {author} {\bibfnamefont {M.~R.}\ \bibnamefont
  {Parsa}},\ }\emph {\bibinfo {title} {Lattice gases with molecular dynamics
  collision operator}},\ \href@noop {} {Ph.D. thesis},\ \bibinfo  {school}
  {North Dakota State University} (\bibinfo {year} {2018})\BibitemShut
  {NoStop}%
\bibitem [{sup()}]{supplemental}%
  \BibitemOpen
  \href@noop {} {}\bibinfo {note} {URL will be inserted by
  publisher}\BibitemShut {NoStop}%
\bibitem [{\citenamefont {Wagner}\ and\ \citenamefont
  {Strand}(2016)}]{wagner2016fluctuating}%
  \BibitemOpen
  \bibfield  {author} {\bibinfo {author} {\bibfnamefont {A.~J.}\ \bibnamefont
  {Wagner}}\ and\ \bibinfo {author} {\bibfnamefont {K.}~\bibnamefont
  {Strand}},\ }\bibfield  {title} {\bibinfo {title} {Fluctuating lattice
  boltzmann method for the diffusion equation},\ }\href@noop {} {\bibfield
  {journal} {\bibinfo  {journal} {Physical Review E}\ }\textbf {\bibinfo
  {volume} {94}},\ \bibinfo {pages} {033302} (\bibinfo {year}
  {2016})}\BibitemShut {NoStop}%
\end{thebibliography}%

\end{document}